\documentclass[twoside,12pt]{article}
\usepackage{graphicx}

\begin{document}

\centerline{{\it Brazilian Journal of Physics} {\bf 33}, 616 (2003)}
\bigskip

\centerline{\Large \bf Corrections to Finite Size Scaling in
Percolation}
\bigskip
P.M.C. de Oliveira$^1$, R.A. N\'obrega$^2$ and D. Stauffer$^3$.

\bigskip
Instituto de F\'\i sica, Universidade Federal Fluminense, av. Litor\^anea
s/n, Boa Viagem, Niter\'oi, Brasil 24210-340

\medskip

\noindent $^1$ pmco@if.uff.br \space  $^2$ rafaella@if.uff.br

\noindent $^3$ stauffer@thp.uni-koeln.de; Visiting from Institute for
Theoretical Physics, Cologne University, D-50923 K\"oln, Euroland

\bigskip

Abstract: A $1/L$-expansion for percolation problems is proposed, where $L$
is the lattice finite length. The square lattice with 27 different sizes $L
= 18, 22, \dots 1594$ is considered. Certain spanning probabilities were
determined by Monte Carlo simulations, as continuous functions of the site
occupation probability $p$. We estimate the critical threshold $p_c$ by
applying the quoted expansion to these data. Also, the universal spanning
probability at $p_c$ for an annulus with aspect ratio $r = 1/2$ is estimated
as $C = 0.876657(45)$.

\section{Introduction}

	In reference \cite{nz} a square lattice is viewed as a torus, thus
without frontiers, where sites are randomly occupied with probability $p$.
{\sl Wrapping} percolation probabilities can be defined within this
geometry, counting configurations which wrap along the horizontal and/or/xor
vertical directions. The probability $R_h$, for instance, counts all
configurations wrapping along the horizontal direction, no matter which
occurs vertically. As a function of $p$, $R_h$ corresponds to a plot like
figure 1 (to be quoted later) for a finite lattice. In the thermodynamic
limit $L \to \infty$, this plot approaches a step function, $R_h = 0$ below,
and $R_h = 1$ above the critical threshold $p_c$. The point $p_L(\tau)$
exemplified on the plot serves as an estimator for $p_c$: by measuring a
sequence of such values, for larger and larger sizes $L_1, L_2 \dots L_N$,
one can extrapolate this sequence for $L \to \infty$. Reference \cite{nz}
presents the figure $p_c = 0.59274621(13)$, the most accurate available
today, obtained from more than $7 \times 10^9$ Monte Carlo samples. Starting
from an empty lattice, filled site by site at random, each sample provides a
single number to the statistics, namely the precise number $n$ of occupied
sites for which the proper wrap (horizontal for $R_h$) appears for the first
time. This computational strategy of filling up the lattice site by site and
storing data onto bell-shaped $n$-histograms is equivalent to that of
\cite{elias}, and similar to early works \cite{domb}. However, the
multi-step strategy introduced in \cite{nz} involves many other components
(see \cite{nz2}). The authors of \cite{nz} assume a $p_L(\tau) - p_c \sim
L^{-(2+1/\nu)}$ dependence, where $\nu = 4/3$ is the correlation length
critical exponent. As $2+1/\nu = 2.75$ is a sufficiently high value, data up
to only $L = 128$ were needed. Also, the authors adopt the value $\tau^* =
0.521058290$, corresponding to that particular wrapping geometry and exactly
known \cite{pinson} as the limiting value for $R_h$ at $p_c$, called
Pinson's number in \cite{zlk}. Outside this value or within other
geometries, the not-so-high exponent $1+1/\nu$ or even the smaller $1/\nu$
were observed \cite{ziff}.

\begin{figure}[hbt]
\begin{center}
\includegraphics[angle=-90,scale=0.45]{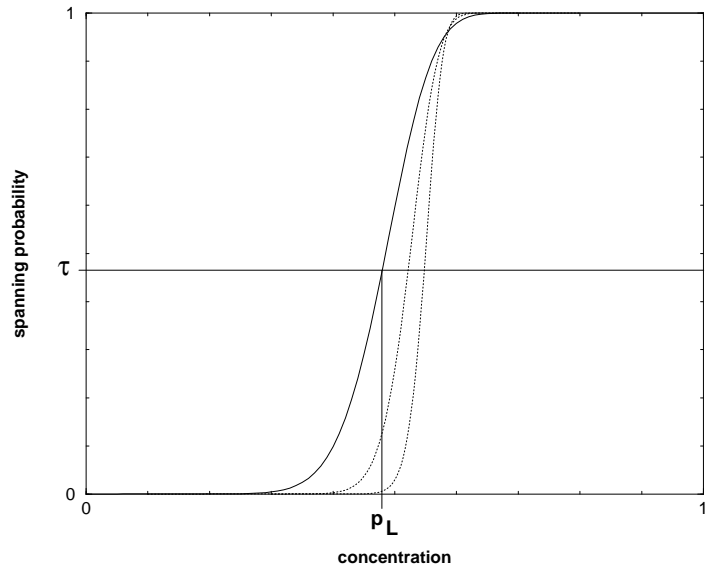}
\end{center}
\caption{Spanning probability function for $L = 18$ (solid line), $38$ and
$74$ (dotted lines). By fixing some value $\tau$, one can find a sequence of
values $p_L(\tau)$ for increasing lattice sizes, approaching the critical
threshold $p_c$.}
\end{figure}

	Here, we propose the mathematical form

\begin{equation}
p_L(\tau) = p_c + {1\over L^{1/\nu}} \Big[ A_0(\tau) + {A_1(\tau)\over L} +
{A_2(\tau)\over L^2} + \dots {A_M(\tau)\over L^M} \Big]\,\,\,\, ,
\end{equation}

\noindent for estimators $p_L$ obtained from quantities like $R_h$, where
the cutoff $M$ is conveniently chosen according to the numerical accuracy
available.

\begin{table}[hbt]
\begin{center}
\begin{tabular}{|l|c|c|}
\hline
$L$ & $p_L(\tau=0.9)$ & samples\\
\hline

  18 & 0.55982808(075) & $10^9$\\
  22 & 0.56196704(122) & $10^9$\\
  26 & 0.56403393(062) & $10^9$\\
  30 & 0.56590384(082) & $10^9$\\
  38 & 0.56902009(083) & $10^9$\\
  46 & 0.57146271(087) & $10^9$\\
  52 & 0.57296244(060) & $10^9$\\
  62 & 0.57500000(066) & $10^9$\\
  74 & 0.57689631(175) & $10^8$\\
  86 & 0.57838778(129) & $10^8$\\
 102 & 0.57993924(157) & $10^8$\\
 118 & 0.58115110(109) & $10^8$\\
 142 & 0.58254835(133) & $5\times 10^7$\\
 166 & 0.58360542(109) & $5\times 10^7$\\
 202 & 0.58479326(182) & $3\times 10^7$\\
 234 & 0.58558821(076) & $3\times 10^7$\\
 282 & 0.58649210(124) & $10^7$\\
 334 & 0.58721227(157) & $10^7$\\
 402 & 0.58791296(095) & $10^7$\\
 478 & 0.58848866(104) & $10^7$\\
 566 & 0.58898678(159) & $6\times 10^6$\\
 674 & 0.58944374(126) & $6\times 10^6$\\
 802 & 0.58984063(146) & $4\times 10^6$\\
 958 & 0.59019986(149) & $4\times 10^6$\\
1126 & 0.59048760(168) & $4\times 10^6$\\
1354 & 0.59077660(121) & $4\times 10^6$\\
1594 & 0.59100202(082) & $4\times 10^6$\\

\hline
\end{tabular}
\caption{Values for $p_L$ obtained for fixed $\tau=0.9$, as an example. A
Chi-square fitting of equation (1) gives $p_c = 0.59274675(88)$, $A_0 =
-0.44204(11)$, $A_1 = 3.275(14)$, etc.}
\end{center}
\end{table}

	We apply this formula to the $L \times L$ square lattice, with a set
of different increasing lengths such that the numbers of sites grow by a
factor of $\sqrt{2}$. Table I shows an example for $\tau = 0.9$, for which
the traditional Chi-square fitting \cite{recipes} gives $p_c =
0.59274675(88)$, in agreement with \cite{nz} although within a larger error
bar. We adopted $M = 4$ in equation (1), compatible with our smallest
lattice size $L = 18$, since $18^{-4.75} = 1 \times 10^{-6}$ still falls
inside our numerical accuracy, whereas the next term $18^{-5.75} = 6 \times
10^{-8}$ would be outside. The quality of this fit can be appreciated by the
so-called goodness-of-fit $Q$ \cite{recipes}, a quantity between 0 and 1.
The fit is considered believable \cite{recipes} for values of $Q > 0.1$. In
our example, we get $Q = 0.857$ for table I. All our data to be discussed
hereafter, for many other values of $\tau$ between $0.5$ and $0.99$, present
the same degree of accuracy, giving credit to our proposal, equation (1). In
spite of these accurate results, one cannot rule out some possible higher
terms deviating from (1), for example that proposed in \cite{hovi}.

\section{Measured quantity}

	First, let's explain the {\sl spanning} probability we adopted
within the torus, instead of the {\sl wrapping} probability \cite{nz}. We
consider two parallel lines distant $L/2$ from each other, on the $L \times
L$ square lattice. For each sample --- again obtained by filling up the
initially empty lattice, site by site at random --- we count the precise
number $n$ of occupied sites for which these lines become connected for the
first time, no matter which occurs around the other direction. This approach
has a big advantage over the {\sl wrapping} probability around the whole
torus: From the same sample we can count $n$ just $L$ times instead of only
once! The parallel lines can be numbered ($i,i+L/2$) for $i = 1, 2, \dots
L/2$ along the horizontal direction, with the same procedure repeated
vertically. Thus, the statistics is multiplied by a factor of $L$. In table
I, for instance, the sampling counting $10^9$ for $L = 18$ corresponds to an
$n$-histogram with $18 \times 10^{9}$ accumulated units (the total area
below the bell-shaped curve). In the same table, for $L = 1594$, the much
smaller sampling counting $4 \times 10^6$ corresponds indeed to almost the
same statistics, i.e. $6 \times 10^{9}$ accumulated units below the curve.
This trick allows us to test a wide range of lattice sizes, and verify the
validity of our proposal (1). The further computational time one needs in
order to implement this trick is negligible: we simply keep in memory the
top and bottom (left and right) extreme lines for each already formed
cluster of neighbouring occupied sites. Thus, for each new included site,
only the last updated cluster must be verified.

	The {\sl spanning} probability function is obtained by superimposing
a lot of step functions, one for each counted $n$, and dividing the result
by the total number ($L \times$sampling counts). An average is then
performed, weighted by $C(L^2,n) p^n (1-p)^{L^2-n}$ where $C$ is a
combinatorial factor, yielding the $p$-continuous curves shown in figure 1.

	The 3-digits error bars shown in table I were obtained by dividing
the whole set of data into, say, $S = 10$ sub-sets, independently
calculating $p_L(\tau)$ for each sub-set. The error bars are the standard
deviation of this distribution divided by $\sqrt{S}$. This last division is
based on the supposition that the whole data is normally distributed. In
order to verify the validity of this approach, we repeated the same
procedure with $S = 20$. Indeed, the error bars are approximately the same,
independent of $S$. For safety, we adopted always the largest between both
error bars so obtained. For intermediate lattice sizes (from $L = 74$ up to
402), we used $S = 10$ and 5, instead of 20 and 10.

	We also simulated larger $L \times L$ lattices for $\tau = 0.5$, up
to $L = 24000$, with free instead of periodic boundary conditions, within a
poorer statistics. The results are also compatible with equation (1).

\section{Data Analysis}

	By fitting data with equation (1), we have a set of $M+2$ parameters
to be determined, namely $p_c, A_0(\tau), A_1(\tau), \dots A_M(\tau)$. The
error bars for these quantities come from the Chi-square fitting procedure
\cite{recipes}, as a consequence of the primary error bars directly measured
for the crude data (that of table I, for instance). Thus, the accuracy
obtained for each parameter, in particular the interesting one $p_c$,
involves a series of accumulated errors. Instead of taking care of $p_c$,
let's turn our attention to $A_0(\tau)$ for a while. Figure 2 shows a plot
of this value as a function of $\tau$.

\begin{figure}[hbt] \begin{center}
\includegraphics[angle=-90,scale=0.5]{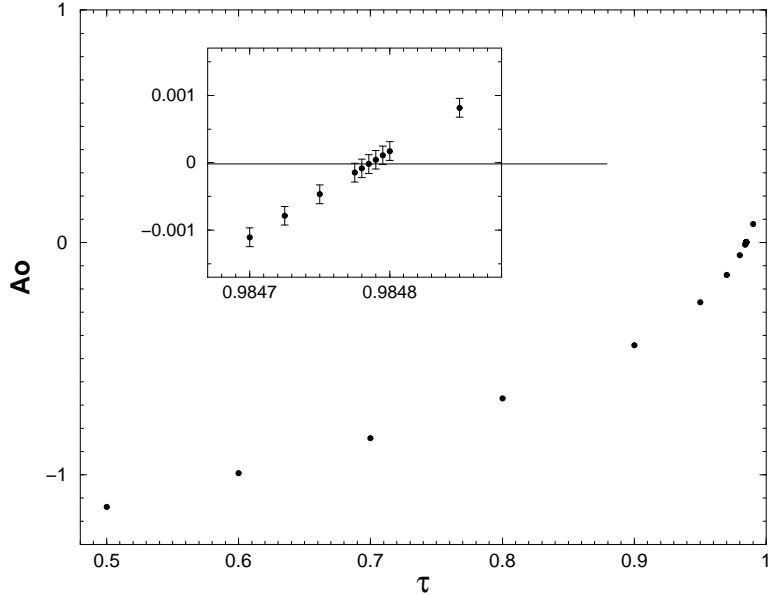} \end{center}
\caption{Expansion parameter $A_0$, equation (1), as a function of $\tau$.
The error bars obtained from the Chi-square fitting appear only within the
smaller scale of the inset, near the special point $\tau^* = 0.984786(11)$
where $A_0$ vanishes. }
\end{figure}

	One can see that $A_0$ vanishes for the particular value $\tau^* =
0.984786(11)$. For the geometry we adopted, this is the equivalent of the
above quoted Pinson's number \cite{pinson,zlk}, i.e. the universal value of
our spanning probability at $p_c$ in the thermodynamic limit. Recently,
Cardy \cite{cardy} tried to calculate this value for an annulus, i.e. an
$L_1 \times L_2$ square lattice where periodic boundary condition is adopted
only along the direction of $L_1$. The quoted universal value corresponds to
the critical spanning probability between the two free-boundary lines, and
depends only on the ratio $r = L_2/L_1$. Unfortunately, within his
theoretical approach, Cardy was forced to leave out spanning configurations
which also wrap along the direction of $L_1$, the periodic boundary.

	Our geometry with the complete $L \times L$ torus divided in two
halves by the lines $i$ and $i + L/2$ corresponds to the $r = 1/2$ Cardy's
geometry counted twice, in parallel. Thus, our $\tau^*$ can be related to
Cardy's number $C$ by $1 - \tau^* = (1 - C)^2$, i.e. $C = 0.876657(45)$.
Compared with Cardy's exact value $C - C_{\rm wrap} = 0.7569977963$
\cite{cardy} we found the contribution $C_{\rm wrap} = 0.119659(45)$ for the
spanning configurations which also wrap around the periodic direction.

	The critical occupation probability $p_c$ can be obtained from our
expansion (1) by fixing any value for $\tau$. The best choice is $\tau =
\tau^*$, for which the leading finite-size term in (1) becomes
$L^{-1-1/\nu}$ instead of $L^{-1/\nu}$. In this case, we get $p_c =
0.59274621(33)$, where the error bar is estimated by the same above-quoted
procedure of dividing the whole data set into $S = 10$ sub-sets. For some
unknown particular reason, the {\sl wrapping} probabilities adopted in
\cite{nz} works better yet, the leading finite-size term in (1) being
$L^{-2-1/\nu}$. In \cite{nz} a numerical evidence for that behaviour is
given, by plotting $\tau_L(p_c) - \tau^*$ against $L$ (assuming some
previously determined value for $p_c$) and verifying a power-law dependence
with an exponent very close to $-2$. Indeed, for our {\sl spanning}
probability, the same plot gives an exponent very close to $-1$.

\section{Conclusion}

	We propose the finite-size expansion (1) for spanning probabilities
in percolation, where $p_L(\tau)$ is defined in figure 1.

	This approach can be used to calculate various critical quantities.
We applied it to a particular geometry within the site percolation problem
on a $L \times L$ square lattice, considered as a torus. Taking two parallel
lines distant $L/2$ from each other, our spanning probability counts
configurations for which these two lines are connected. The universal value
$\tau^* = 0.984786(11)$ and the critical occupation $p_c = 0.59274621(33)$
are obtained.

	As a by-product, we propose the universal value $C = 0.876657(45)$
for the critical spanning probability within a $L \times L/2$ annulus, i.e.
a square lattice with periodic (free) boundaries along the direction of $L$
($L/2$). This probability corresponds to all configurations for which the
frontiers separated by $L/2$ are connected. Cardy \cite{cardy} determined
the exact figure $C - C_{\rm wrap} = 0.7569977963$, where $C_{\rm wrap}$
corresponds to spanning configurations which also wrap along the direction
of $L$, discounted in his approach.

	The whole computer time for obtaining these data was 20 thousand
hours, on a dozen computers, typically powered by an Athlon 1GHz processor.

\section{Acknowledgements}

	Six months ago, the first author presented a preliminary version of
this work in Campos do Jord\~ao, S\~ao Paulo, Brazil, in honour of the
60$^{\rm th}$ birthday of professor Silvio Salinas. Now, some other articles
are presented in honour of the 60$^{\rm th}$ birthday of the last author. In
contrast, the younger authors dedicated this work to the senior professor
Silvio Salinas.

	We are indebted to Carlos Tomei, Antonio Brito Serbeto, John Cardy
and Robert Ziff for helpful discussions, and Brazilian agencies CNPq and
FAPERJ for support.

\end{document}